\documentclass{article}
\usepackage{amsmath}
\usepackage{cite}
\usepackage{graphicx}
\usepackage{dcolumn}

\begin{document}

\title{On the screened Kratzer potential and its variants}
\author{Francisco M. Fern\'{a}ndez\thanks{%
fernande@quimica.unlp.edu.ar} \\
INIFTA, DQT, Sucursal 4, C. C. 16, \\
1900 La Plata, Argentina}
\maketitle

\begin{abstract}
We argue that several potentials proposed recently for the analysis of the
vibrational-rotational spectra of diatomic molecules and their thermodynamic
properties exhibit a flaw. One can easily show that the parameters $D_e $
and $r_e$ in those potentials are not the dissociation energy and
equilibrium bond length, respectively, as the proposers believe. We show how
to overcome the mistake in a simple and quite general way.
\end{abstract}

\section{Introduction}

\label{sec:intro}

Several years ago, Kratzer\cite{K20} proposed a potential for the analysis
of the spectra of diatomic molecules and some time later Fues\cite{F26}
solved the Schr\"{o}dinger equation with this potential. For this reason,
the potential is commonly called Kratzer potential or Kratzer-Fues
potential. A few years later, typical textbooks on spectroscopy\cite{H50}
scarcely resort to the Kratzer potential and today no spectroscopist would
take it seriously. However, several authors have recently shown some
interest in the Kratzer potential and even proposed some variants\cite
{IOSR19, IEAORA20, PPR20, PPR21, IORAEAAH21, RIEO21, EIOORLK21, IOSRR22,
BKSBC23}, like the screened Kratzer potential\cite{IOSR19, IEAORA20,
EIOORLK21}, the screened cosine Kratzer potential\cite{PPR20}, the
Hulten-screened Kratzer potential\cite{PPR21}, the improved screened Kratzer
potential\cite{IORAEAAH21}, the improved Kratzer potential\cite{RIEO21}, the
shifted screened Kratzer potential\cite{IOSRR22} and the harmonic plus
screened Kratzer potential\cite{BKSBC23}.

The purpose of this note is the discussion of all those molecular
potentials. In section~\ref{sec:potentials} we analyze the
potentials just mentioned, in section~\ref{sec:correct_pot} we
show how to generate them correctly and in
section~\ref{sec:conclusions} we summarize the main results and
draw conclusions.

\section{The modified Kratzer potentials}

\label{sec:potentials}

Before discussing the potentials we review a relevant feature of a potential
$V(r)$ for a diatomic molecule. According to any textbook on spectroscopy
the equilibrium bond length $r_{e}$ and the dissociation energy $D_{e}$ are
given by\cite{H50}
\begin{equation}
\left. \frac{dV(r)}{dr}\right|
_{r=r_{e}}=0,\;D_{e}=\lim\limits_{r\rightarrow \infty }V(r)-V(r_{e}),
\label{eq:re,De,V}
\end{equation}
to which we should add $V^{\prime \prime }(r_{e})>0$ because the stationary
point at $r=r_{e}$ should be a minimum.

The Kratzer potential can be written in several equivalent forms; in what
follows we choose the expression used in most of the papers mentioned in the
introduction:
\begin{equation}
V_{K}(r)=-2D_{e}\left( \frac{r_{e}}{r}-\frac{r_{e}^{2}}{2r^{2}}\right) .
\label{eq:V_K}
\end{equation}
Note that $V_{K}(r)$ satisfies equations (\ref{eq:re,De,V}).

The first variant of the Kratzer potential is the Screened Kratzer potential
\begin{equation}
V_{SK}(r)=-2D_{e}\left( \frac{r_{e}}{r}-\frac{r_{e}^{2}}{2r^{2}}\right)
e^{-\alpha r},  \label{eq:V_SK}
\end{equation}
proposed by Ikot et al\cite{IOSR19} and used also in other papers\cite
{IEAORA20, EIOORLK21} for some physical applications. In this case, $\alpha
\geq 0$ is a screening parameter. One can \textit{easily verify} that $%
V_{SK}(r)$ does not satisfy equations (\ref{eq:re,De,V}):
\begin{equation}
V_{SK}^{\prime }\left( r_{e}\right) =\alpha D_{e}e^{-\alpha
r_{e}},\;V_{SK}\left( r\rightarrow \infty \right) -V_{SK}\left( r_{e}\right)
=D_{e}e^{-\alpha r_{e}}\leq D_{e}.  \label{eq:V_SK_test}
\end{equation}
This variant of the Kratzer potential only satisfies equations (\ref
{eq:re,De,V}) in the trivial case $\alpha =0$. Note that $r=r_{e}$ is not at
the minimum of the potential but to the right of it. Consequently, the
parameters $D_{e}$ and $r_{e}$ in equation (\ref{eq:V_SK}) are not the
dissociation energy and equilibrium bond length, respectively. For this
reason, all the physical applications based on such assumption\cite
{IOSR19,IEAORA20, EIOORLK21} are of doubtful utility.

Purohit et al\cite{PPR20} proposed the screened cosine Kratzer potential
\begin{equation}
V_{SCK}(r)=-2D_{e}\left( \frac{r_{e}}{r}-\frac{r_{e}^{2}}{2r^{2}}\right)
e^{-\alpha r}\cosh (\delta \alpha r),  \label{eq:V_SCK}
\end{equation}
where $\delta $ is another screening parameter. The authors are not clear
about the suitable values of $\delta $ and here we assume that $-1\leq
\delta \leq 1$ so that $e^{-\alpha r}\cosh (\delta \alpha r)\rightarrow 0$
when $r\rightarrow \infty $. Purohit et al chose the trivial value $\delta =0
$ and also $\delta =1$. A straightforward calculation leads to
\begin{eqnarray}
V_{SCK}^{\prime }\left( r_{e}\right)  &=&\alpha D_{e}e^{-\alpha r_{e}}\left[
\cosh \left( \alpha \delta r_{e}\right) -\delta \sinh \left( \alpha \delta
r_{e}\right) \right] ,\;  \nonumber \\
V_{K}\left( r\rightarrow \infty \right) -V\left( r_{e}\right)
&=&D_{e}e^{-\alpha r_{e}}\cosh \left( \alpha \delta r_{e}\right) .
\label{eq:VSCK_test}
\end{eqnarray}
We appreciate that the parameters $D_{e}$ and $r_{e}$ are not de
dissociation energy and equilibrium bond length, respectively. Once again we
conclude that all the physical results and conclusions derived from this
assumption may not be correct\cite{PPR20}.

Purohit et al\cite{PPR21} also proposed the Hulth\'{e}n-screened cosine
Kratzer potential
\begin{equation}
V_{HSCK}(r)=-\frac{V_{0}e^{-\alpha r}}{1+e^{-\alpha r}}-2D_{e}\left( \frac{%
r_{e}}{r}-\frac{r_{e}^{2}}{2r^{2}}\right) e^{-\delta \alpha r}\cosh (\delta
\lambda \alpha r),  \label{eq:V_HSCK}
\end{equation}
where $\lambda $ is another screening parameter. Once again, the authors are
unclear about the values of the screening constants. Here, we assume that $%
\alpha \geq 0$, $\delta \geq 0$ and $-1\leq \lambda \leq 1$. The authors
chose $\lambda =0,1/2,1$ in their applications. This potential does not
satisfy equations (\ref{eq:re,De,V}) as shown by
\begin{eqnarray}
&&V_{HSCK}^{\prime }\left( r_{e}\right) =\alpha \delta D_{e}e^{-\alpha
\delta r_{e}}\left[ \cosh \left( \alpha \delta \lambda r_{e}\right) -\lambda
\sinh \left( \alpha \delta \lambda r_{e}\right) \right] +\frac{\alpha
V_{0}e^{\alpha r_{e}}}{\left( 1+e^{\alpha r_{e}}\right) ^{2}},  \nonumber \\
&&V_{HSCK}\left( r\rightarrow \infty \right) -V_{HSCK}\left( r_{e}\right)
=D_{e}e^{-\alpha \delta r_{e}}\cosh \left( \alpha \delta \lambda
r_{e}\right) +\frac{V_{0}}{1+e^{\alpha r_{e}}}.  \label{eq:V_HSCK_test}
\end{eqnarray}
As in the preceding examples, $D_{e}$ and $r_{e}$ are not the molecular
parameters just mentioned.

Ikot et al\cite{IORAEAAH21} also proposed the improved screened Kratzer
potential (also known as improved Kratzer potential\cite{RIEO21})
\begin{equation}
V_{ISK}(r)=-2D_{e}\left( \frac{r_{e}}{r}-\frac{r_{e}^{2}}{2r^{2}}\right)
\left[ e^{-\frac{\alpha +\delta }{2}r}\cosh \left( \frac{\alpha +\delta }{2}%
r\right) +\tau \right] ,  \label{eq:V_ISK}
\end{equation}
where $\tau $ is a control parameter with values $-1,$ $0$ and $1$. This
potential can be simplified as
\begin{equation}
V_{ISK}(r)=-2D_{e}\left( \frac{r_{e}}{r}-\frac{r_{e}^{2}}{2r^{2}}\right)
\left[ \frac{e^{-(\alpha +\delta )r}}{2}+\tau +\frac{1}{2}\right] .
\end{equation}
Despite its improvement, this potential does not satisfy equations (\ref
{eq:re,De,V}) because
\begin{eqnarray}
V_{ISK}^{\prime }\left( r_{e}\right)  &=&\frac{(\alpha +\delta )D_{e}e^{-%
\frac{\alpha +\delta }{2}r_{e}}}{2}\left[ \cosh \left( \frac{\alpha +\delta
}{2}r_{e}\right) -\sinh \left( \frac{\alpha +\delta }{2}r_{e}\right) \right]
,  \nonumber \\
V_{ISK}(r &\rightarrow &\infty )-V_{ISK}\left( r_{e}\right) =D_{e}\left[ e^{-%
\frac{\alpha +\delta }{2}r_{e}}\cosh \left( \frac{\alpha +\delta }{2}%
r_{e}\right) +\tau \right] .  \label{eq:V_ISK_test}
\end{eqnarray}
It is clear that $D_{e}$ and $r_{e}$ do not have their intended meaning.

Ibrahim et al\cite{IOSRR22} invented the shifted screened Kratzer potential
\begin{equation}
V_{SSK}(r)=-2D_{e}\left( \frac{r_{e}}{r}-\frac{r_{e}^{2}}{2r^{2}}\right)
\left( 2\lambda +\gamma e^{-\alpha r}\right) ,  \label{eq:V_SSK}
\end{equation}
where $\lambda $ and $\gamma $ are shifting parameters. The expressions
\begin{equation}
V_{SSK}^{\prime }\left( r_{e}\right) =\alpha \gamma D_{e}e^{-\alpha
r_{e}},\;V_{SK}\left( r\rightarrow \infty \right) -V_{SK}\left( r_{e}\right)
=D_{e}\left( \gamma e^{-\alpha r_{e}}+2\lambda \right) ,
\label{eq:V_SSK_test}
\end{equation}
undoubtedly show that $r_{e}$ and $D_{e}$ are not the equilibrium bond
length and dissociation energy, respectively.

Finally, we mention the harmonic plus screened Kratzer potential of Bansal
et al\cite{BKSBC23}
\begin{equation}
V_{HSK}(r)=-2D_{e}\left( \frac{r_{e}}{r}-\frac{r_{e}^{2}}{2r^{2}}\right)
e^{-\alpha r}+cr^{2}.  \label{eq:V_HSK}
\end{equation}
Since there are no bound states when $c<0$ we only consider $c\geq 0$. For $%
c=0$ we have the screened Kratzer potential discussed above and for $c>0$
(the novelty of this proposal) this potential does not predict dissociation
because $V_{HSK}(r\rightarrow \infty )=\infty $. Not only there is no
dissociation energy but $r_{e}$ is not the equilibrium bond length because
\begin{equation}
V_{HSK}^{\prime }\left( r_{e}\right) =\alpha D_{e}e^{-\alpha r_{e}}+2cr_{e}.
\label{eqV_HSK_test}
\end{equation}

In conclusion, we have shown that all the potentials described above suffer
from the same flaw: in all of them $r_{e}$ and $D_{e}$ are not the
equilibrium bond length and the dissociation energy, respectively. In the
physical applications the authors substituted the experimental values of the
equilibrium bond length and dissociation energy into those model parameters.
For this reason the vibrational-rotational energies that they calculated and
showed in several tables appear to be of scarce utility. Note that they did
not attempt to compare their theoretical results with experimental data. In
the next section we show how to modify the potentials discussed above in
such a way that the model parameters $r_{e}$ and $D_{e}$ have the correct
meaning.

\section{The correct form of the potentials}

\label{sec:correct_pot}

Most of the potentials described in the preceding section are particular
cases of
\begin{equation}
V_{G}(r)=\left( \frac{a}{r}+\frac{b}{r^{2}}\right) f(r),
\end{equation}
where we assume that $f(r\rightarrow \infty )=0$. If we substitute this
expression into equations (\ref{eq:re,De,V}) we obtain a system of two
equations with two unknowns: $a$ and $b$. Upon solving such system of
equations we obtain the desired potential. A straightforward calculation
shows that
\begin{equation}
V_{G}(r)=-\frac{2D_{e}f(r)}{f\left( r_{e}\right) }\left[ \frac{r_{e}}{r}-%
\frac{r_{e}^{2}}{2r^{2}}+\frac{r_{e}f^{\prime }\left( r_{e}\right) }{%
2f\left( r_{e}\right) }\left( \frac{r_{e}^{2}}{r^{2}}-\frac{r_{e}}{r}\right)
\right] .  \label{eq:V_G}
\end{equation}
We appreciate that this expression yields the Kratzer potential when $%
f(r)\equiv 1$. When $f(r)=e^{-\alpha r}$ we obtain the correct form of the
shifted Kratzer potential
\begin{equation}
V_{SK}(r)=D_{e}\left[ \frac{\left( \alpha r_{e}+1\right) r_{e}^{2}}{r^{2}}-%
\frac{\left( \alpha r_{e}+2\right) r_{e}}{r}\right] e^{-\alpha \left(
r-r_{e}\right) }.  \label{eq:VSK_correct}
\end{equation}
One can easily verify that this form of the potential already satisfies
equations (\ref{eq:re,De,V}). We can proceed in the same way with the other
potentials but we do not deem it necessary.

In closing this section, we mention that the Hulth\'{e}n-screened cosine
Kratzer potential is a particular case of
\begin{equation}
V_{G2}(r)=\left( \frac{a}{r}+\frac{b}{r^{2}}\right) f(r)+g(r),
\end{equation}
where, for convenience, we choose $g(r\rightarrow \infty )=0$. We can easily
obtain suitable expressions for $a$ and $b$ as in the preceding case so that
the resulting potential will satisfy equations (\ref{eq:re,De,V}).

\section{Further comments and conclusions}

\label{sec:conclusions}

In section~\ref{sec:potentials} we showed that several potentials proposed
recently for the analysis of the rotational-vibrational spectra of diatomic
molecules and their thermodynamic properties exhibit a serious flaw. The
model parameters $r_{e}$ and $D_{e}$ in those potentials are not the
equilibrium bond length and dissociation energy, respectively, as the
proposers believed. The authors inserted experimental values for those model
parameters obtaining potentials that are unsuitable for the intended
physical application. For this reason, the results obtained by those authors
are of doubtful utility.

In section~\ref{sec:correct_pot} we solved the problem in a simple way. We
thus arrived at a general expression that gives the correct form of those
potentials in which $r_{e}$ and $D_{e}$ are the molecular parameters
mentioned above. However, it is worth noting that the calculation of
vibrational-rotational energies by means of an empirical potential is of no
utility whatsoever\cite{H50}. Any serious spectroscopist would fit the
eigenvalues of the model to the molecular spectrum in order to obtain the
desired model parameters. The quality of the model is given by the square
deviation of the fit.

\end{document}